\documentclass[a4paper,11pt]{article}
\usepackage{wrapfig}
\usepackage{pos}
\usepackage{caption}
\usepackage{lipsum} 

\title{From PeV to TeV: Astrophysical Neutrinos with Contained Vertices in 10 years of IceCube Data}

\ShortTitle{Medium Energy Starting Events}

\author{The IceCube Collaboration \\{\normalsize \normalfont(a complete list of authors can be found at the end of the proceedings)}\\}

\emailAdd{vedant.basu@icecube.wisc.edu}
\emailAdd{aswathi.balagopalv@icecube.wisc.edu}

\abstract{

The IceCube Neutrino Observatory is a cubic-kilometer Cherenkov detector at the South Pole, designed to study neutrinos of astrophysical origin. We present an analysis of the Medium Energy Starting Events (MESE) sample, a veto-based event selection that selects neutrinos and efficiently rejects a background of cosmic ray-induced muons This is an extension of the High Energy Starting Event (HESE) analysis, which established the existence of high-energy neutrinos of astrophysical origin. The HESE sample is consistent with a single power law spectrum with best-fit index $2.87^{+0.20}_{-0.19}$, which is softer than complementary IceCube measurements of the astrophysical neutrino spectrum. While HESE is sensitive to neutrinos above 60 TeV, MESE improves the sensitivity to lower energies, down to 1 TeV. In this analysis we use an improved understanding of atmospheric backgrounds in the astrophysical neutrino sample via more accurate modeling of the detector self-veto. A previous measurement with a 2-year MESE dataset had indicated the presence of a possible 30 TeV excess. With 10 years of data, we have a larger sample size to investigate this excess. We will use this event selection to measure the cosmic neutrino energy spectrum over a wide energy range. The flavor ratio of astrophysical neutrinos will also be discussed.

\vspace{4mm}
{\bfseries Corresponding authors:}
Vedant Basu$^{1*}$, Aswathi Balagopal V.$^{1}$\\
{$^{1}$ \itshape University of Wisconsin-Madison}\\[4mm]
$^*$ Presenter

\ConferenceLogo{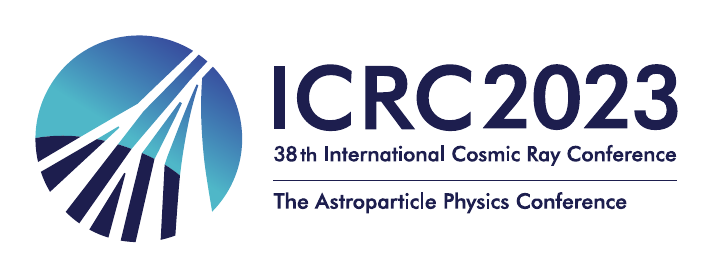}

\FullConference{The 38th International Cosmic Ray Conference (ICRC2023)\\ 26 July -- 3 August, 2023\\ Nagoya, Japan}
}

\begin{document}

\maketitle

\section{Introduction}\label{sec1}
The IceCube Neutrino Observatory is a cubic-kilometer astronomical research facility located at the South Pole. The detector array uses the Antarctic glacier as a Cherenkov medium to detect high-energy astrophysical neutrinos via their interactions with nucleons in the ice.  Reconstruction of the incident neutrino events
 relies on the collection of Cherenkov photons emitted by charged particles produced as result of the neutrino interactions using optical detectors (DOMs) embedded in ice \cite{Aartsen:2016nxy}. The events detected within the detector typically follow three morphologies: cascades, tracks, and double cascades. Cascades are shower-like events produced via charged-current (CC) interaction of electron neutrinos ($\nu_e$) and neutral-current (NC) interactions of all three neutrino flavours. Tracks are mostly generated when muon neutrinos ($\nu_\mu$) undergo CC interactions, while double cascades are a distinct feature of tau neutrino ($\nu_\tau$) CC interactions. Starting events form a special class of events detected with IceCube. Here, the interaction of the neutrino with the ice occurs within the detector volume and therefore a majority of the initial hadronic interaction is visible. The advantages of such 'starting event' samples is that they are sensitive to all neutrino flavours, and to events from all directions.

 \section{Event Selection}
 \label{sec:event_selection}
The data sample focuses on starting events with energies greater than 1 TeV, and is known as the Medium Energy Starting Events (MESE) selection. It extends the High Energy Starting Events (HESE) selection \cite{Abbasi:2020jmh}, the dataset developed for the discovery of astrophysical neutrinos with IceCube, to lower energies. The wider energy range, increased statistics, and improved track energy resolution will help resolve finer spectral features, making it possible to test flux models down to 1 TeV. While HESE utilizes a single outer veto layer to reject higher energy atmospheric muons, MESE uses an additional sequence of vetoes to reject lower energy atmospheric muons and thereby achieves increased sensitivity to astrophysical neutrinos at lower energies. The MESE selection discussed here is an updated version of the historical MESE sample using two years of IceCube data \cite{2015PhRvD..91b2001A}. 
The first stage of the event selection utilizes the outer layer of the detector array to veto incoming muons. This veto region includes all the outer strings, a layer at the top of the detector with 90~m thickness, a 120~m thick layer around the dust-dominated region close to the center of the detector, and a single layer of DOMs at the bottom of the array. Bright HESE events, which deposit at least 6000 photoelectrons (PE) of charge in the detector volume, are permitted up to 3 hits in this veto layer. Events with charge less than 6000~PE are required to deposit no hits in the veto region, to help reject the lower energy atmospheric muons. All HESE-tagged events are retained in the final sample, whereas MESE events must undergo further cuts to reject atmospheric backgrounds. The outer-layer veto successfully rejects over 99.99\% of the bright muons entering the detector.

The events that survive the outer-layer veto are dominated by low-energy muons which escape the veto layer before undergoing a large stochastic energy loss within the detector. To reject these dim muons, one looks for hits associated with various randomly sampled track hypotheses. Events with greater than 2~PE of incoming  charge associated with a track hypothesis are rejected by this stage of the veto. The events that are retained at this stage are classified as tracks or cascades with the help of a neural-network-based classifier \cite{Glauch:2019}.

A large fraction of low-energy muon background passes the second stage of the veto. For dim events it is necessary to ensure that the events start further inside the detector, as any potential muon-induced track would have a greater probability of depositing charge along its way into the detector. A charge, and direction-dependent fiducial volume cut is therefore used to suppress the muon background, which yields the final sample. This fiducial-volume cut is applied separately for track-classified and cascade-classified events, and retains more cascades than tracks within the final event sample. Since atmospheric muons are downgoing, the fiducial volume cut rejects more events from the Southern sky than the Northern sky due to its zenith-angle dependency. All events that pass this stage of the veto are retained in the final event selection.
\begin{figure}[h]
    \centering
    \includegraphics[width=0.7\textwidth]{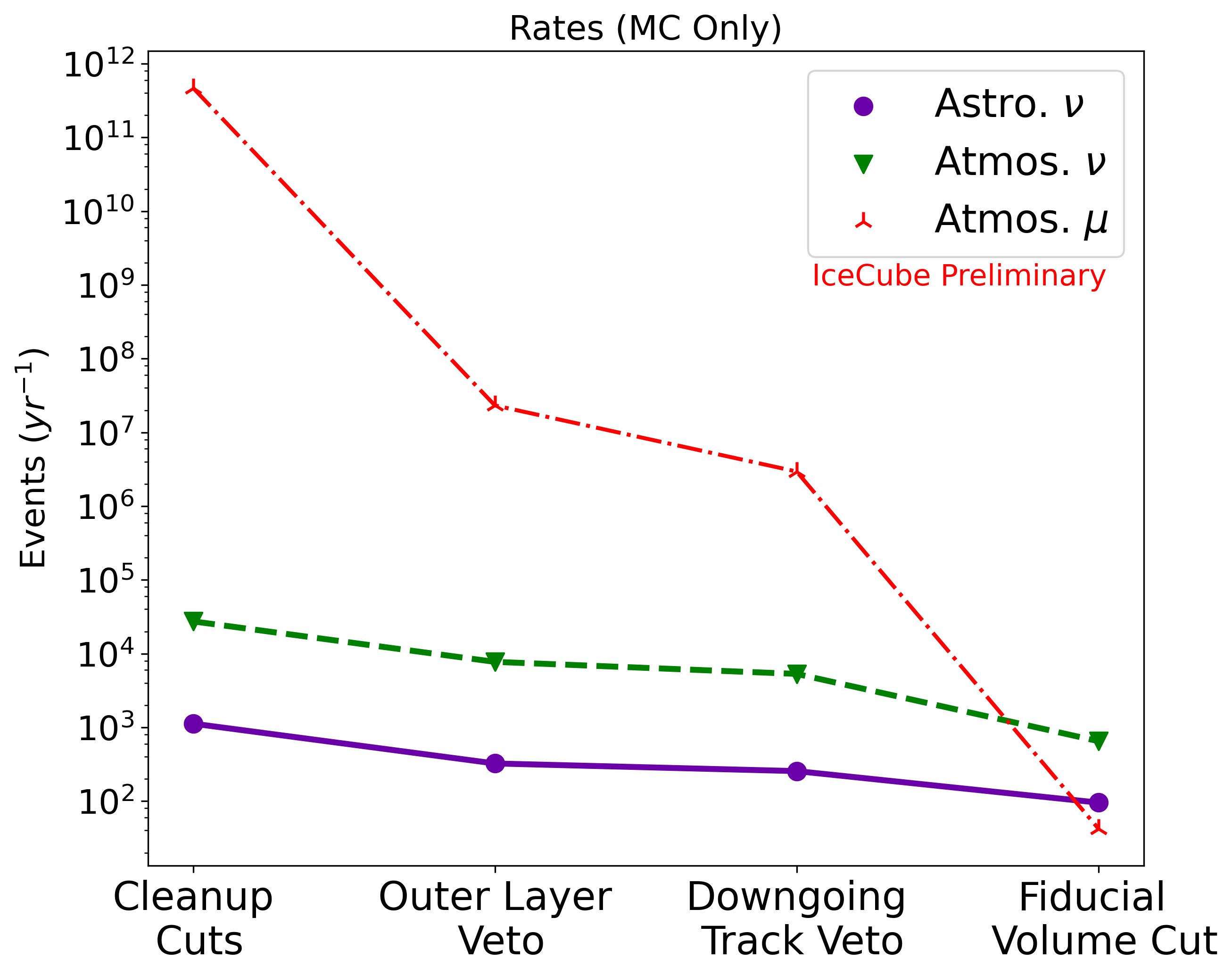}
    \caption{We depict the expected rates of signal (astrophysical neutrinos) and background (atmospheric neutrinos and muons) after each level of the event selection, derived from MC simulation. Cleanup cuts include cuts on the minimum charge deposited, and the number of modules which see hits. This removes low energy events which are difficult to reconstruct accurately. We see that the muon background is suppressed by ten orders of magnitude at the final level, improving our signal/background ratio
}
    \label{fig:RateComparison}
\end{figure}

\subsection{Expected Rates}
The event sample has a large fraction of events from the Northern sky, especially atmospheric neutrinos at lower energies. In the Southern sky, a suppression of atmospheric neutrinos is observed due to the self-veto effect\cite{Arguelles:2018awr}. This effect occurs when atmospheric neutrinos are vetoed due to accompanying muons from the same air shower. Therefore, the Southern sky sees a higher fraction of astrophysical neutrinos, although the rate is 10 times lower than the Northern sky.
The events from the Southern sky constrain the prompt atmospheric flux with the help of the self veto, while the Northern sky events contribute statistical power in constraining the conventional flux. Together these strongly constrain the various components of the overall measurement.  Fig. \ref{fig:RateComparison} illustrates the suppression in background rates with each cut level, while Table~\ref{tab:rate_breakdown} shows us the expected final level rates, from simulations.


\begin{table}
\caption{A breakdown of the expected final level rates from simulation per year by flux component, and by hemisphere. The neutrino and muon simulation is used to model the astrophysical and atmospheric fluxes, with the normalization set to 1 for the atmospheric conventional, atmospheric prompt, and the atmospheric muon fluxes drawn from the "GaisserH4a" composition model\cite{gaisser2012spectrum} with the "SIBYLL2.3c" Hadronic Interaction model \cite{riehn2019hadronic}. The astrophysical flux is drawn from the 2-year MESE result \cite{2015PhRvD..91b2001A}.}
\centering

\label{tab:rate_breakdown}
\begin{tabular}{|c|c|c|c|}
    \hline
    Flux Component  & Expected Rates ($yr^{-1}$) & Northern Sky ($yr^{-1}$) & Southern Sky ($yr^{-1}$) \\
    \hline
    Astro $\nu$  & 95.48 & 70.32& 25.16  \\
    \hline
     Atm. $\nu$  & 650.10& 610.89 & 39.21\\
    \hline
     Atm. $\mu$  & 42.38 &21.62&20.76  \\
    \hline
\end{tabular}
\end{table}

\section{Measurement of the Astrophysical Neutrino Flux}
\subsection{Flux Components}
The dataset will contain astrophysical neutrinos, atmospheric conventional and prompt neutrinos, and atmospheric muons. We model the atmospheric background events with a normalization scaling factor multiplying the expectation from a flux model. The astrophysical neutrino flux is modeled here assuming a single power law flux, and a 1:1:1 neutrino flavor ratio.
\subsection{Analysis Method}
The diffuse astrophysical spectrum is measured using a multi-component forward folding binned likelihood analysis. We divide the events by morphology into track-like and cascade-like events, based on a Deep Neural Net classifier\cite{Glauch:2019}. The observables used for the fit are the event reconstructed energy and cosine zenith angle\cite{yuan2023detecting}. The energy range considered for binning spans from 1 TeV to 10 PeV. Since cascade events are reconstructed with better energy resolution, they are divided into 22 bins, while tracks are binned more coarsely into 13 bins. Similarly, the reconstructed cosine zenith angles are divided into 10 bins for the binned-likelihood test. The form of likelihood used for the analysis is an effective likelihood\cite{arguelles2019binned}, which accounts for finite Monte-Carlo (MC) simulations of the various components of the fit. In the limit of large MC statistics, this effective likelihood converges to a Poissonian form.

The main component of the fit we are interested in measuring is the astrophysical flux normalization and the spectral index. The atmospheric flux components of the likelihood fit include the normalization of the conventional neutrino flux, the normalization of the muon component and the normalization of the atmospheric neutrino flux induced by the decay of the charm component of air showers. We also account for the uncertainties associated with the air-shower modeling using a parameterization from Barr et al\cite{2006PhRvD..74i4009B}, and a gradient term that allows interpolation between the Gaisser-H4a  cosmic ray composition model\cite{2012APh....35..801G}, which is treated as the baseline model, and the GST \cite{2013FrPhy...8..748G} models which show the strongest differences in the atmospheric-neutrino spectra resulting from their respective primary cosmic-ray predictions. Several detector systematics are also taken into account for the fit. Uncertainties arise due to the properties of the ice like its absorption coefficient, scattering coefficient and anisotropies within the bulk ice. The refreezing of the ice surrounding the optical modules during deployment also adds a systematic effect, since it scatters light more than the bulk ice. The uncertainty of the optical efficiency of the DOMs also acts as a systematic. These detector systematics are included within the fit as perturbations to the baseline of their respective values. This is done within the MC using the Snowstorm method, which is described in detail in \cite{2019JCAP...10..048A}. 

The atmospheric self-veto effect, described earlier, is also an important systematic to account for, requiring the accurate characterisation of the detector response to accompanying muons. For estimating its effect, we injected muons of varying energies to the final level neutrino sample and estimated the number of events that get vetoed due to this accompanying muon. With this method, we obtain probabilities of the neutrino getting vetoed for each energy, zenith, and entering at each depth of the detector. While we estimate this effect using MC, there could be uncertainties that arise due to mismodeling of this effect. Therefore, we also account for this as a nuisance parameter within our fit.

\subsection{Sensitivity}
\begin{figure}[h]
    \centering
\includegraphics[width=0.65\linewidth]{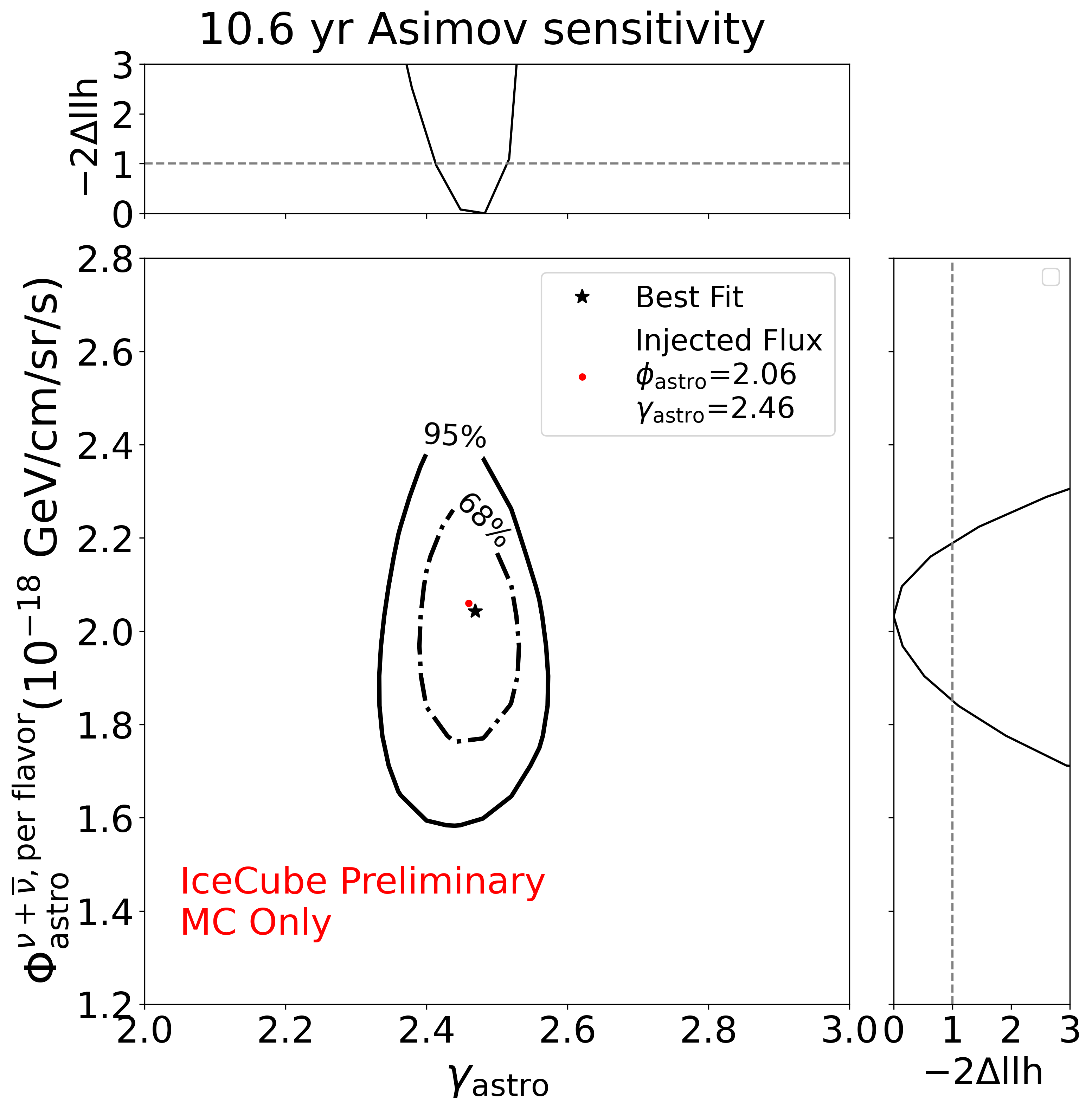}
    \caption{Two-dimensional Asimov profile likelihood scan of the astrophysical flux assuming a Single Power Law (SPL) flux model. We perform a grid scan over each pair of points representing the astrophysical normalization $\phi_{astro}$ and spectral index $\gamma_{astro}$.  The injected flux is derived from the the 2-year MESE result. Assuming Wilk's theorem holds, we compute the 68\% and 95\% confidence intervals depicted. We incorporate detector systematic effects following the SnowStorm method\cite{2019JCAP...10..048A}, to model the effect of DOM Efficiency, Hole and Bulk Ice systematics as nuisance parameters. Additional systematics affecting the atmospheric neutrino flux are also modelled, including the self-veto effect}
    \label{fig:sensitivities}
\end{figure}

We show a two-dimensional Asimov profile likelihood scan and calculate the expected 68\% and 95\% confidence intervals for sensitivity to the astrophysical flux parameters assuming a single power law, seen in Figure \ref{fig:sensitivities}. We assume 10.6 years of IceCube-86 data here, corresponding to the seasons from 2011-2021. We inject the astrophysical flux model from the 2-year MESE result  (($\phi_{\mathrm{astro}},\gamma_{\mathrm{astro}})$= (2.06, 2.46)) \cite{2015PhRvD..91b2001A}. As the prompt and astrophysical neutrino fluxes follow similar spectral shapes, there is a degeneracy between the normalization factors which expands the uncertainties on $\phi_{\mathrm{astro}}$. The degeneracy between the astrophysical and the prompt neutrino fluxes is broken by the atmospheric self-veto, which suppresses the prompt atmospheric neutrino flux prediction in the Southern sky, distinguishing the prompt flux from the astrophysical flux, which is expected to be isotropic. 
\section{Measurement of the Astrophysical Flavour Ratio}
The MESE dataset contains astrophysical neutrinos of all flavours from the entire sky. This can help in measuring the flavour ratio of astrophysical neutrinos with this sample. As mentioned in Section \ref{sec:event_selection}, the cascade and track events are classified as such as a part of the event-selection procedure. We can also additionally identify the double-cascade events produced by tau neutrinos to make a measurement of the relative fractions of $\nu_e$, $\nu_\mu$ and $\nu_\tau$ events that contribute to the overall astrophysical flux.
\subsection{Selection of Double cascades}
The final level events from the MESE event selection are used to additional classify double-cascade events. For this, we use a likelihood-based regression, used for the HESE flavour measurement \cite{2022EPJC...82.1031I}, which estimates the probability that the event also includes a decaying tau. The MESE events, reconstructed with this routine, are further examined to select the tau neutrino events. For this, we  use the energy ratio (ER) and energy confinement (EC) distributions. We define them as $\mathrm{ER} = \frac{E1-E2}{E1+E2}$  and  $\mathrm{EC} = \frac{E1,C+E2,C}{E_{\mathrm{Tot}}}$ where $E1$ and $E2$ are the energies of the first and second cascade respectively, $E1,C$ and $E2,C$ represent the total deposited energy within 40~m distance of each cascade, and $E_{\mathrm{Tot}}$ is the total deposited energy of the event. $E_{\mathrm{Tot}} = E1,C + E2,C$ is true only for double cascades.
By examining the background $\nu_e$ and $\nu_\mu$ distributions, we have selected the events that pass the condition of EC$\geq$0.99 
and -0.98 $\leq$ ER $\leq$ 0.3 as tau-neutrino events. Additionally we require that each cascade should have at least an energy of 1 TeV. This formalism, the same as that described in \citep{2022EPJC...82.1031I}, works the best for higher energy events. Therefore, we use only events with energy $>$ 30 TeV as the finally selected tau-neutrino event candidates. These selected events have a purity of $\approx 70\%$ of tau-neutrino events. A majority of these events are HESE-classified events due to the high-energy criterion. 
\subsection{Analysis Method}
We use the three types of events identified by MESE: cascades, tracks, and double cascades to conduct a binned-likelihood analysis to measure the astrophysical flavour ratio. The major components for this method are the energy and zenith histograms of each event-topology type. The cascades-energy and zenith histograms are dominated by electron neutrino events. There is significant contamination from NC interactions of tau and muon flavoured neutrinos too. A large fraction of the tau neutrino CC events also contaminate the cascades topology. The tracks topology is dominated by muon neutrino events through their CC interactions. There is 17\% contamination of tau neutrinos that decay into muons in this channel. The double cascades channel, on the other hand, is dominated by tau-neutrinos. There is some amount of contamination from misclassified $\nu_e$ and $\nu_\mu$ events also in this channel. However, the relatively high purity of $\nu_e$ events in cascades, $\nu_\mu$ events in tracks, and $\nu_\tau$ events in double cascades provides the strength to this analysis in making a flavour measurement. 

\begin{figure}
    \centering
    \vspace{-3mm}
    \includegraphics[width=0.75\linewidth]{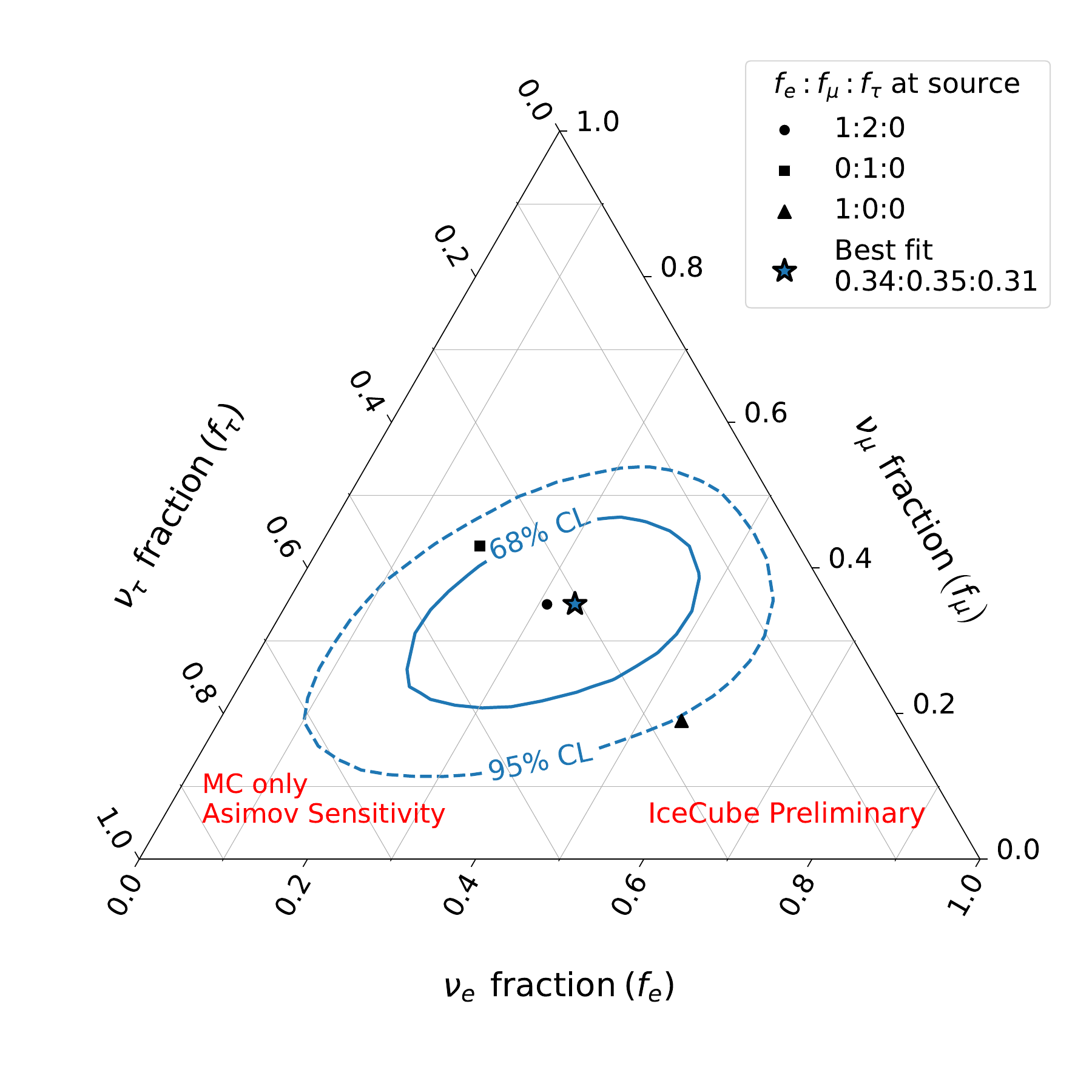}
     \captionsetup{skip=1pt} 
     \vspace{-3mm}
    \caption{The Asimov sensitivity of the flavor composition using the MESE event sample is shown, with 68\% and 95\% confidence intervals assuming 11.3 years of livetime. We inject the astrophysical flux model from the 2-year MESE result (SPL with normalization = 2.06, spectral index = 2.46) \cite{2015PhRvD..91b2001A}, assuming a $\nu_e$:$\nu_\mu$:$\nu_\tau$ ratio of 1:1:1. Also shown are the expected flavor ratios for various source scenarios of the flavour ratio, after taking oscillation effects into account.}
    \label{fig:flav}
\end{figure}

We use the same fit parameters used in the fit for the astrophysical flux measurement, and the systematics are also treated in the same manner. The astrophysical spectral index and normalization also enter as nuisance parameters for the flavour-ratio measurement. We have two additional parameters which account for the fraction of $\nu_e$ ($f_e$) and fraction of $\nu_\tau$ ($f_\tau$) events in the fit. We additionally keep the constraint that $f_e + f_\tau + f_\mu = 1$, where $f_\mu$ represents the fraction of $\nu_\mu$. 
Figure \ref{fig:flav} shows the sensitivity of this analysis to measure the flavour ratio of astrophysical events using 11.3 years of starting events detected by IceCube. The sensitivity projection is determined from a profile likelihood scan in the $f_e$ and $f_\tau$ dimensions and the confidence intervals are determined assuming Wilk's theorem. The analysis mainly depends on the double cascades events with energy $>$ 30 TeV to constrain the contours along the $f_\tau$ axis. The presence of cascade and track events from 1 TeV and above gives additional leverage to the analysis to measure the flavour ratios due to the increased availability of statistics. Additionally, the deep-neural network based cascades/tracks classifier performs well in identifying these topologies which acts as an advantage for this measurement.
\section{Conclusion}
An overview of a veto-based event selection sensitive to the diffuse astrophysical neutrino spectrum from TeV-PeV scales has been presented. We utilize events with vertices contained within the IceCube detector, granting us sensitivity to neutrinos of all flavors from any direction. As an extension of the existing HESE sample to lower energies, and due to the increase in atmospheric background, MESE relies on additional cuts and an updated treatment of systematics to improve sensitivity. The larger sample size and wide energy range helps resolve fine spectral features down to 1 TeV. In addition to the measurement of the diffuse flux, we also present a method to study the flavor ratio of the astrophysical neutrino sample, using a classifier to identify tau neutrino events. We have verified the expected sensitivities with simulations and are proceeding towards a measurement of the entire dataset.





\bibliographystyle{ICRC}
\bibliography{references}

%

\clearpage

\section*{Full Author List: IceCube Collaboration}

\scriptsize
\noindent
R. Abbasi$^{17}$,
M. Ackermann$^{63}$,
J. Adams$^{18}$,
S. K. Agarwalla$^{40,\: 64}$,
J. A. Aguilar$^{12}$,
M. Ahlers$^{22}$,
J.M. Alameddine$^{23}$,
N. M. Amin$^{44}$,
K. Andeen$^{42}$,
G. Anton$^{26}$
% ...
\\
\\
% ...
$^{12}$ Universit{\'e} Libre de Bruxelles, Science Faculty CP230, B-1050 Brussels, Belgium \\
% ...
$^{17}$ Department of Physics, Loyola University Chicago, Chicago, IL 60660, USA \\
$^{18}$ Dept. of Physics and Astronomy, University of Canterbury, Private Bag 4800, Christchurch, New Zealand \\
% ...
$^{22}$ Niels Bohr Institute, University of Copenhagen, DK-2100 Copenhagen, Denmark \\
$^{23}$ Dept. of Physics, TU Dortmund University, D-44221 Dortmund, Germany \\
% ...
$^{26}$ Erlangen Centre for Astroparticle Physics, Friedrich-Alexander-Universit{\"a}t Erlangen-N{\"u}rnberg, D-91058 Erlangen, Germany \\
% ...
$^{40}$ Dept. of Physics and Wisconsin IceCube Particle Astrophysics Center, University of Wisconsin{\textendash}Madison, Madison, WI 53706, USA \\
% ...
$^{42}$ Department of Physics, Marquette University, Milwaukee, WI, 53201, USA \\
% ...
$^{44}$ Bartol Research Institute and Dept. of Physics and Astronomy, University of Delaware, Newark, DE 19716, USA \\
% ...
$^{63}$ Deutsches Elektronen-Synchrotron DESY, Platanenallee 6, 15738 Zeuthen, Germany  \\
$^{64}$ Institute of Physics, Sachivalaya Marg, Sainik School Post, Bhubaneswar 751005, India \\
% ...

\subsection*{Acknowledgements}

\noindent
The authors gratefully acknowledge the support from the following agencies and institutions:
USA {\textendash} U.S. National Science Foundation-Office of Polar Programs,
U.S. National Science Foundation-Physics Division,
U.S. National Science Foundation-EPSCoR,
Wisconsin Alumni Research Foundation,
Center for High Throughput Computing (CHTC) at the University of Wisconsin{\textendash}Madison,
Open Science Grid (OSG),
Advanced Cyberinfrastructure Coordination Ecosystem: Services {\&} Support (ACCESS),
Frontera computing project at the Texas Advanced Computing Center,
U.S. Department of Energy-National Energy Research Scientific Computing Center,
Particle astrophysics research computing center at the University of Maryland,
Institute for Cyber-Enabled Research at Michigan State University,
and Astroparticle physics computational facility at Marquette University;
Belgium {\textendash} Funds for Scientific Research (FRS-FNRS and FWO),
FWO Odysseus and Big Science programmes,
and Belgian Federal Science Policy Office (Belspo);
Germany {\textendash} Bundesministerium f{\"u}r Bildung und Forschung (BMBF),
Deutsche Forschungsgemeinschaft (DFG),
Helmholtz Alliance for Astroparticle Physics (HAP),
Initiative and Networking Fund of the Helmholtz Association,
Deutsches Elektronen Synchrotron (DESY),
and High Performance Computing cluster of the RWTH Aachen;
Sweden {\textendash} Swedish Research Council,
Swedish Polar Research Secretariat,
Swedish National Infrastructure for Computing (SNIC),
and Knut and Alice Wallenberg Foundation;
European Union {\textendash} EGI Advanced Computing for research;
Australia {\textendash} Australian Research Council;
Canada {\textendash} Natural Sciences and Engineering Research Council of Canada,
Calcul Qu{\'e}bec, Compute Ontario, Canada Foundation for Innovation, WestGrid, and Compute Canada;
Denmark {\textendash} Villum Fonden, Carlsberg Foundation, and European Commission;
New Zealand {\textendash} Marsden Fund;
Japan {\textendash} Japan Society for Promotion of Science (JSPS)
and Institute for Global Prominent Research (IGPR) of Chiba University;
Korea {\textendash} National Research Foundation of Korea (NRF);
Switzerland {\textendash} Swiss National Science Foundation (SNSF);
United Kingdom {\textendash} Department of Physics, University of Oxford.

\end{document}